\documentclass[times,english]{llncs}
\usepackage{url}
\usepackage[english]{babel}

\usepackage{amssymb}
\usepackage{amsmath}
\usepackage{graphicx}
\usepackage{subfigure}
\usepackage{fancyvrb}
\usepackage[textwidth=3cm,textsize=footnotesize]{todonotes}
\usepackage{algpseudocode}
\usepackage{listings}
\usepackage{tlatex}
\usepackage{color}
\usepackage{subfigure}
\definecolor{boxshade}{gray}{0.7}
\setboolean{shading}{true}

\newcommand{\figscale}{.65}
\newcommand{\uml}[1]{\text{{\sffamily\setbox0=\hbox{X}\fontsize{1.25\ht0}{\baselineskip}\selectfont
      #1}}} 
\newcommand{\stereotype}[1]{\flqq{\uml{#1}}\frqq}
\newcommand{\spa}{SPA}

\newcommand{\ta}{TLA$^{+}$}
\newcommand{\var}[1]{\ensuremath{\mathit{#1}}}
\newcommand{\true}{\ensuremath{\textsc{true}}}
\newcommand{\false}{\ensuremath{\textsc{false}}}

\begin{document}
\title{Specifying and Model Checking Workflows of Single Page Applications with \ta}
\author{Gefei Zhang}

\institute{Hochschule f\"ur Technik und Wirtschaft Berlin\\
\texttt{gefei.zhang@htw-berlin.de}
}

\maketitle

\begin{abstract}
  Single Page Applications (\spa{}s) are different than hypertext-based web applications
  in that their workflow is not defined by explicit links, but rather implicitly
  by changes of their widgets' states.  The workflow may hence be hard to track.
  We present an approach to specifying and model checking \spa{}s
  with \ta. Our approach makes it easier to
  document and to track the workflow of \spa{}s and to find potential design
  flaws.
\end{abstract}

\section{Introduction}
\label{sec:intro}

Single Page Applications (\spa{}s) are web applications that ``interact
with the user by dynamically rewriting the current page rather than loading
entire new pages from a server''~\cite{wikipedia:spa}.  Compared with
hypertext-based web applications,
\spa{}s do not have an explicit navigation structure.  Instead, their
workflow is controlled implicitly by the states of their control elements
(widgets).  For example, widgets may appear or disappear to provide different
information to the user, the information provided by a widget may change over
time, and widgets may be disabled to disallow the user to perform a certain
action.  

While formal specification and validation of the navigation structure has been
proved valuable in the area of hypertext-based web applications,
see~\cite{alzahrani:phd:2015}, analysis of 
workflows of \spa{}s has been little investigated.  In this paper, we
present an approach to specifying and model checking~\cite{clarke_et_al:2018}
the workflow of \spa{}s with \ta. 

\ta~\cite{lamport:2003} is a formal specification
language based on Temporal Logic of Actions.  A \ta{} module consists of
variables and actions. The valuation of the variables determines the current
state of the system, the actions define possible transitions between the states.
\ta{} also provides tools to model check or to theorem proof~\cite{gallier:2018}
  the specification.
Of the two methods, model checking is easier to use since its fully automatic.
This paper focuses on model checking only. 

We use variables to model user observable properties of widgets and actions
to model user actions. The mapping is straight forward and thus practice
friendly.  Our approach helps to understand the requirements for the \spa{}
better, and model checking helps to find potential design errors. 

The reminder of this paper is organized as follows: In the following
Sect.~\ref{sec:tla} we give a brief introduction to \ta. In
Sect.~\ref{sec:specification} we illustrate our approach by a simple example.
We show examples of model checking the specification in
Sect.~\ref{sec:model-checking}. Related work is discussed in
Sect.~\ref{sec:related}, before in Sect.~\ref{sec:conclusions} we draw
conclusions and sketch some future work. 

\section{\ta}
\label{sec:tla}

\ta{} is a formal language for system specification based on Temporal Logic of
Actions. It is used to model temporal behavior of software systems. The models
are amenable to formal verification by model checking or theorem proving. In the
following, we give a brief introduction to \ta{} by an example, see
Fig.~\ref{fig:tla:example}. The source code of this example can be downloaded
from \url{https://bitbucket.org/gefei/tlaweb-models/}.

\begin{figure}[!h]
\begin{tla}
------------------------------- MODULE clock -------------------------------
EXTENDS Naturals
VARIABLES hr, period

vars == <<hr, period>>

Init == /\ hr \in 1..12 
        /\ period \in {"am", "pm"}
Next == /\ hr' = IF hr = 12 THEN 1 ELSE hr + 1 
        /\ period' = IF hr = 11 THEN
                        IF period = "am" THEN "pm" 
                        ELSE "am"
                     ELSE period  
                     
Spec == Init /\ [][Next]_vars /\ WF_vars(Next)        
                 
Invariant == [](hr \in (1..12))
Liveness  == []<>(hr = 1)
=============================================================================
\end{tla}
\begin{tlatex}
\@x{}\moduleLeftDash\@xx{ {\MODULE} clock}\moduleRightDash\@xx{}%
\@x{ {\EXTENDS} Naturals}%
\@x{ {\VARIABLES} hr ,\, period}%
\@pvspace{8.0pt}%
\@x{ vars\@s{2.10} \.{\defeq} {\langle} hr ,\, period {\rangle}}%
\@pvspace{8.0pt}%
\@x{ Init\@s{4.12} \.{\defeq} \.{\land} hr \.{\in} 1 \.{\dotdot} 12}%
\@x{\@s{39.83} \.{\land} period \.{\in} \{\@w{am} ,\,\@w{pm} \}}%
 \@x{ Next \.{\defeq} \.{\land} hr \.{'} \.{=} {\IF} hr \.{=} 12 \.{\THEN} 1
 \.{\ELSE} hr \.{+} 1}%
\@x{\@s{39.83} \.{\land} period \.{'} \.{=} {\IF} hr \.{=} 11 \.{\THEN}}%
\@x{\@s{106.46} {\IF} period \.{=}\@w{am} \.{\THEN}\@w{pm}}%
\@x{\@s{106.46} \.{\ELSE}\@w{am}}%
\@x{\@s{94.30} \.{\ELSE} period}%
\@pvspace{8.0pt}%
 \@x{ Spec\@s{1.46} \.{\defeq} Init \.{\land} {\Box} [ Next ]_{ vars}
 \.{\land} {\WF}_{ vars} ( Next )}%
\@pvspace{8.0pt}%
\@x{ Invariant \.{\defeq} {\Box} ( hr \.{\in} ( 1 \.{\dotdot} 12 ) )}%
\@x{ Liveness\@s{3.71} \.{\defeq} {\Box} {\Diamond} ( hr \.{=} 1 )}%
\@x{}\bottombar\@xx{}%
\end{tlatex}
\caption{\ta{} specification: clock}
  \label{fig:tla:example}
\end{figure}

Our module \var{clock} models a 12-hour-clock.  It defines two variables to
hold the current state of the clock: \var{hr} stores the time, and \var{period}
stores whether it is am or pm.  It is common practice in \ta{} to define a tuple
$vars$ to refer to the variables as a whole.  The formula
\var{Init} defines in the form of conjunction two conditions to be fulfilled in
the clock's initial state: \var{hr} may be any 
number between 1 and 12, and \var{period} may be either \uml{am} or \uml{pm}.

\var{Next} is an \emph{action}. In \ta, actions define how the system state changes
over time, that is, when state change is allowed, and what the succeeding state
is. In the clock example, \var{Next} does not need any precondition, because the
clock is always allowed to proceed to the next hour. The postcondition is
defined in prime variables: \var{hr'} defines the value of \var{hr} in the next
state, and \var{period'} defines the vaule of \var{period} in the next state.
Our action \var{Next} therefore defines declaratively how the clock proceeds to
he next hour: if the current value of \var{hr} is 12, then its new value is 1,
otherwise it is the current value of \var{hr} incremented by 1; If (it is 11 am now) the current
value of \var{hr} is 11 and \var{period} is \uml{am}, then in the next state (12
pm) \var{period} should be \uml{pm}, and if it is 11 pm now, then in the next state
\var{period} should be \uml{am}.

\var{Spec} is the system specification.  It combines the initial state and
state transition rules, and states that the rule \var{Next} is \emph{weakly fair}.
We ignore the precise definition of fairness; the interested reader is referred
to~\cite{kroeger-merz:2008}.  For now, let it suffice to know by declaring
\var{Next} as fair, we make sure that the system does not stutter, that is, our
clock will not stop.

The temporal operator $\square$ means ``always'', that is, the succeeding
formula is true in every state of the system. With its help, it is easy to
define invariants. In our example, the temporal formula \var{Invariant} states
that it is always true that the value of \var{hr} is a number between 1 and 12.
The temporal operation $\diamond$ means ``eventually'', that is,  the succeeding
formula will be eventually true.  The formula \var{Liveness} uses a combination
of the two operators and states that it is always true that sometimes \var{hr}
will be 1. This combination is widely used in temporal logic. A property of the
form ``it is always true that sometimes something will be true'', or ``something
good will eventually happen'', is called a \emph{liveness} property.

TLC confirms that the two properties of our module are true.

\section{Specifying \spa{}s with \ta}
\label{sec:specification}

Due to its action-based nature, \ta{} is well-suited to specifying interactive
systems like \spa{}s.  We illustrate this by an example. The source code of our
model can also be downloaded from
\url{https://bitbucket.org/gefei/tlaweb-models/}.

\begin{figure}[!h]
\begin{center}
\subfigure[Sketch]{
  \raisebox{8mm}{\includegraphics[scale=\figscale]{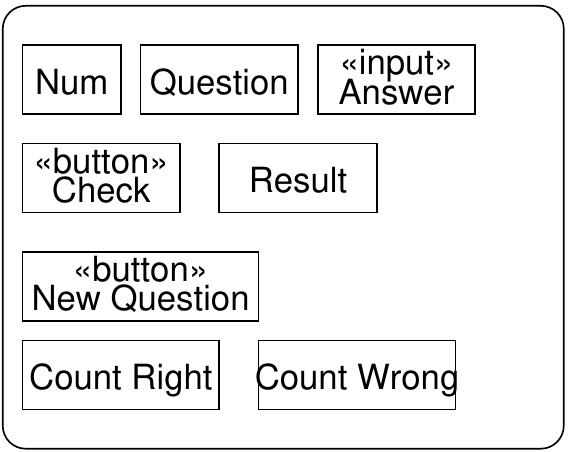}}
 \label{fig:sketch}
}
\subfigure[Sample run]{
  {\includegraphics[scale=.35]{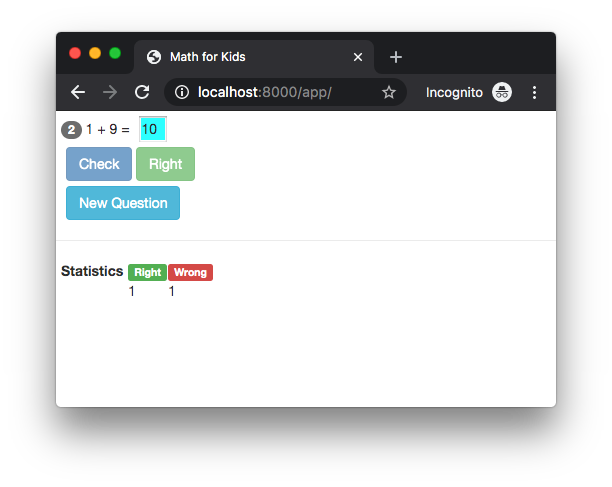}}
  \label{fig:screen}
}
\end{center}
 \caption{Example: math training}
 \label{fig:example}
\end{figure}

Figure~\ref{fig:example} shows a simple \spa{} which provides math exercises
to kids.  A high-level design of the GUI is given in Fig.~\ref{fig:sketch}: the
application should display a
question (\uml{Question}) and its running number (\uml{Num}), wait for the user
to input their answer (\uml{Answer}), and, when the user clicks 
button \uml{Check}, show them if the answer is right or wrong (\uml{Result})
and update a simple 
statistic of the total numbers of right and wrong answers so far (\uml{Count
Right} and \uml{Count Wrong}).  Furthermore, the user should be able to press
button \uml{New Question} to get a new question presented. 
Notice that in Fig.~\ref{fig:sketch} we use \stereotype{input} to identify an
input field and \stereotype{button} to identify buttons. These are the widgets
where the user can interact with the system.
For a better understanding, Fig.~\ref{fig:screen} provides a sample run of the
application where the user has just checked their second answer (which is right)
and has scored one right and one wrong so far.

Suppose it is required that
the user can only click the \uml{Check} button after they have inputted an
answer, and only require a new question after they have finished (checked) the
current one.  Further, we suppose the total number of questions is limited.

\subsection{Constants and Variables}
The \ta{} module modeling this application is given in Fig.~\ref{fig:tla:math}.

First, we need a constant \var{max\_num\_q} to store the maximum number of
questions to present to the user. Then we define a variable for each
\emph{observable} property of the widgets: 
\begin{itemize}
  \item we use a variable \var{num} to store the number of the current question
  \item the field for the user to input their answer may be enabled or disabled,
    we introduce a variable \var{input\_enabled} to model it,
  \item the buttons \uml{Check} and \uml{New Question} may be either enabled or
    disabled, we introduce variables \var{check\_enabled} and
    \var{new\_question\_enabled} to model
    them,
  \item we introduce a variable \var{result} to model the result of checking
    the user's answer to the current question.
  \item we need two variables \var{count\_right} and \var{count\_wrong} to
    hold the numbers of right and wrong answers so far, 
\end{itemize}

\begin{figure}
\begin{tla}
----------------------------- MODULE math -----------------------------
EXTENDS Naturals

CONSTANT max_num_q

VARIABLE num, count_right, count_wrong, result,
         input_enabled, check_enabled, new_question_enabled

vars == <<num, count_right, count_wrong, result,
          input_enabled, check_enabled, new_question_enabled>>

Init == /\ num = 1
        /\ count_right = 0
        /\ count_wrong = 0
        /\ input_enabled = TRUE
        /\ check_enabled = FALSE
        /\ new_question_enabled = FALSE
        /\ result = ""
        
Input_Answer == /\ input_enabled = TRUE
                /\ input_enabled' = FALSE
                /\ check_enabled' = TRUE
                /\ UNCHANGED <<num, count_right, count_wrong, new_question_enabled, result>>
                
Check == /\ check_enabled = TRUE
         /\ check_enabled' = FALSE
         /\ new_question_enabled' = TRUE
         /\ \E answer_correct \in {TRUE, FALSE}:
            IF answer_correct = TRUE THEN
               /\ count_right' = count_right + 1
               /\ result' = "Right"
               /\ UNCHANGED count_wrong                     
            ELSE /\ count_wrong' = count_wrong + 1
                 /\ UNCHANGED count_right
                 /\ result' = "Wrong" 
         /\ UNCHANGED <<num, input_enabled>>
                
New_Question == /\ num < max_num_q
                /\ new_question_enabled = TRUE
                /\ new_question_enabled' = FALSE
                /\ num' = num + 1
                /\ input_enabled' = TRUE
                /\ result' = ""
                /\ UNCHANGED <<count_right, count_wrong, check_enabled>>

Terminating == /\ num = max_num_q 
               /\ UNCHANGED vars

Next ==    \/ Input_Answer 
           \/ Check 
           \/ New_Question
           \/ Terminating

Spec == Init  /\ [][Next]_vars /\ WF_vars(Next)
====
\end{tla}
\begin{tlatex}
\@x{}\moduleLeftDash\@xx{ {\MODULE} math}\moduleRightDash\@xx{}%
\@x{ {\EXTENDS} Naturals}%
\@pvspace{8.0pt}%
\@x{ {\CONSTANT} max\_num\_q}%
\@pvspace{8.0pt}%
\@x{ {\VARIABLE} num ,\, count\_right ,\, count\_wrong ,\, result ,\,}%
\@x{\@s{46.84} input\_enabled ,\, check\_enabled ,\, new\_question\_enabled}%
\@pvspace{8.0pt}%
 \@x{ vars \.{\defeq} {\langle} num ,\, count\_right ,\, count\_wrong ,\,
 result ,\,}%
 \@x{\@s{41.61} input\_enabled ,\, check\_enabled ,\, new\_question\_enabled
 {\rangle}}%
\@pvspace{8.0pt}%
\@x{ Init\@s{2.02} \.{\defeq} \.{\land} num \.{=} 1}%
\@x{\@s{37.72} \.{\land} count\_right\@s{5.30} \.{=} 0}%
\@x{\@s{37.72} \.{\land} count\_wrong \.{=} 0}%
\@x{\@s{37.72} \.{\land} input\_enabled\@s{0.63} \.{=} {\TRUE}}%
\@x{\@s{37.72} \.{\land} check\_enabled \.{=} {\FALSE}}%
\@x{\@s{37.72} \.{\land} new\_question\_enabled \.{=} {\FALSE}}%
\@x{\@s{37.72} \.{\land} result \.{=}\@w{}}%
\@pvspace{8.0pt}%
\@x{ Input\_Answer \.{\defeq} \.{\land} input\_enabled \.{=} {\TRUE}}%
\@x{\@s{81.04} \.{\land} input\_enabled \.{'}\@s{0.63} \.{=} {\FALSE}}%
\@x{\@s{81.04} \.{\land} check\_enabled \.{'} \.{=} {\TRUE}}%
 \@x{\@s{81.04} \.{\land} {\UNCHANGED} {\langle} num ,\, count\_right ,\,
 count\_wrong ,\, new\_question\_enabled ,\, result {\rangle}}%
\@pvspace{8.0pt}%
\@x{ Check \.{\defeq} \.{\land} check\_enabled \.{=} {\TRUE}}%
\@x{\@s{45.52} \.{\land} check\_enabled \.{'} \.{=} {\FALSE}}%
\@x{\@s{45.52} \.{\land} new\_question\_enabled \.{'} \.{=} {\TRUE}}%
 \@x{\@s{45.52} \.{\land} \E\, answer\_correct\@s{4.93} \.{\in} \{ {\TRUE} ,\,
 {\FALSE} \} \.{:}}%
\@x{\@s{56.63} {\IF} answer\_correct \.{=} {\TRUE} \.{\THEN}}%
\@x{\@s{68.78} \.{\land} count\_right \.{'} \.{=} count\_right \.{+} 1}%
\@x{\@s{68.78} \.{\land} result \.{'} \.{=}\@w{Right}}%
\@x{\@s{68.78} \.{\land} {\UNCHANGED} count\_wrong}%
 \@x{\@s{56.63} \.{\ELSE} \.{\land} count\_wrong \.{'} \.{=} count\_wrong
 \.{+} 1}%
\@x{\@s{87.94} \.{\land} {\UNCHANGED} count\_right}%
\@x{\@s{87.94} \.{\land} result \.{'} \.{=}\@w{Wrong}}%
 \@x{\@s{45.52} \.{\land} {\UNCHANGED} {\langle} num ,\, input\_enabled
 {\rangle}}%
\@pvspace{8.0pt}%
\@x{ New\_Question \.{\defeq} \.{\land} num \.{<} max\_num\_q}%
\@x{\@s{82.75} \.{\land} new\_question\_enabled \.{=} {\TRUE}}%
\@x{\@s{82.75} \.{\land} new\_question\_enabled \.{'} \.{=} {\FALSE}}%
\@x{\@s{82.75} \.{\land} num \.{'} \.{=} num \.{+} 1}%
\@x{\@s{82.75} \.{\land} input\_enabled \.{'} \.{=} {\TRUE}}%
\@x{\@s{82.75} \.{\land} result \.{'} \.{=}\@w{}}%
 \@x{\@s{82.75} \.{\land} {\UNCHANGED} {\langle} count\_right ,\, count\_wrong
 ,\, check\_enabled {\rangle}}%
\@pvspace{16.0pt}%
\@x{ Terminating \.{\defeq} \.{\land} num \.{=} max\_num\_q}%
\@x{\@s{73.56} \.{\land} {\UNCHANGED} vars}%
\@pvspace{8.0pt}%
\@x{ Next \.{\defeq}\@s{12.29} \.{\lor} Input\_Answer}%
\@x{\@s{52.13} \.{\lor} Check}%
\@x{\@s{52.13} \.{\lor} New\_Question}%
\@x{\@s{52.13} \.{\lor} Terminating}%
\@pvspace{8.0pt}%
 \@x{ Spec\@s{1.46} \.{\defeq} Init\@s{4.1} \.{\land} {\Box} [ Next ]_{ vars}
 \.{\land} {\WF}_{ vars} ( Next )}%
\@x{}\bottombar\@xx{}%
\end{tlatex}

\caption{\ta{} specification: math}
\label{fig:tla:math}
\end{figure}

\subsection{Actions}
We define a TLA+ action for each action the user may take.  In our sample
application, the user may input an answer to the current problem, check if the
answer is correct, or get a new problem.

The action \var{Input\_Answer} models the user action of inputting an answer.
The user is only allowed to do this when the input widget is enabled
(\var{input\_enabled} = \true).  After this action, the only thing we need to
change is that the user should be allowed to check their answer now
(\var{change\_enabled'} = \true).  All other variable should hold their old
values.

The action \var{Check} models the user action of clicking the button
\uml{Check} and letting the system check if the answer is correct.  The only
precondition is that \var{check\_enabled} must be \true.  After this action,
this variable should be set to \false{} to disable the action. Also,
\var{new\_question\_enabled} is set to \true{} to allow the user to get another
question.  The correctness of the user input is simulated by
\var{answer\_correct}, and the variables \var{result}{} and \var{count\_right}{}
updated accordingly.

The action \var{New\_Question} models the user action of getting a new
question. The precondition is $\var{new\_question\_enabled} = \true$ and
$\var{num} < \var{max\_num\_q}$, since we produce  at most \var{max\_num\_q}
questions. The action sets \var{new\_question\_enabled} to \false, and
\var{input\_enabled} to \true{} to allow the user to enter their answer.

The state of the application after initialization is modeled in the
action \var{Init}. Since the applications shows the user a first question upon
initialization, the number of the current question is 1; The user is allowed to
input an answer, but not to check the correctness of the answer nor to get a new
question yet. 

When the maximum number of questions has been reached ($\var{num} =
\var{max\_num\_q}$), then the system is
terminated. We model this by the action \var{Terminating}, which simply says
that the variables do not change any more.
Finally, \var{Next} comprises the three user actions and \var{Terminating}, and
\var{Spec} defines the overall system behavior. As in the 12-hour clock example,
we declare \var{Next} to be weakly fair to prevent the system from stuttering.

\section{Model Checking}
\label{sec:model-checking}

Specifying our \spa{} with \ta{} allows us to verify some properties of our model
formally. This way, \ta{} is helpful for finding potential design flaws.  In the
following we will use TLC, \ta's model checker, to do this.

\subsubsection*{Reachability.} We first examine the reachability of all
questions, that is, for every number $x \in \{1..\var{max\_num\_q}\}$ it holds
that sometimes the number of the current question is $x$.
This property can be defined as follows:

\medskip
\begin{center}
\begin{tla}
Reachability == \A x \in 1..max_num_q: <>(num = x)
\end{tla}
\begin{tlatex}
 \@x{ Reachability \.{\defeq} \A\, x \.{\in} 1 \.{\dotdot} max\_num\_q \.{:}
 {\Diamond} ( num \.{=} x )}%
\end{tlatex}
\end{center}

\noindent
Running TLC confirms that this property is true.

\subsubsection*{Liveness.} We also examine that if the user is able to input
an answer, then sometimes they will also be able to get a new question. In order
to express this property, we need the temporal operator $\leadsto$. In \ta,
property $F \leadsto G$ means whenever $F$ is true, then eventually $G$ will be
true.  In the following, we check that in our module, whenever \var{input\_enabled} is \true,
then sometimes \var{new\_question\_enabled} will also be \true.

\medskip
\begin{center}
  \begin{tla}
    Liveness == input_enabled ~> new_question_enabled
  \end{tla}
\begin{tlatex}
 \@x{\@s{16.4} Liveness \.{\defeq} input\_enabled \.{\leadsto}
 new\_question\_enabled}%
\end{tlatex}
\end{center}

\medskip
\noindent
TLC also confirms that this property holds.

\subsubsection*{Invariant.} Since every answer is checked, it seems obvious that
the number of the current question
should always equal the sum of the numbers of right and wrong answers so far. We
formulate this property as follows:

\begin{center}
\begin{tla}
  Invariant == [](num = count_right + count_wrong)           \* Does not hold!
\end{tla}
\begin{tlatex}
 \@x{\@s{8.2} Invariant \.{\defeq} {\Box} ( num \.{=} count\_right \.{+}
 count\_wrong )\@s{41.0}}%
\@y{%
  Does not hold!
}%
\@xx{}%
\end{tlatex}
\end{center}

\medskip
\noindent
However, TLC reports that this property does not hold. The reason is that
$\var{count\_right} + \var{count\_wrong}$ yields the number of all
questions that have been \emph{checked}, but there may be one 
question which is presented to the user but not checked yet.  Therefore 
$num\_q = count\_right + count\_wrong$ only holds after the action \var{Check}
(and before \var{New\_Question}). That is, it only holds when \var{result} is
not empty.  The following invariant can therefore be verified by TLC:

\begin{center}
\begin{tla}
  Invariant == [](result = "" \/ num = count_right + count_wrong)   \* Correct 
\end{tla}
\begin{tlatex}
 \@x{\@s{8.2} Invariant \.{\defeq} {\Box} ( result \.{=}\@w{} \.{\lor} num
 \.{=} count\_right \.{+} count\_wrong )\@s{8.2}}%
\@y{%
  Correct 
}%
\@xx{}%
\end{tlatex}
\end{center}

\subsubsection*{Deadlock.}
Even applications as simple as the example may be easily erroneous. For example, 
if we made a ``hit-by-one'' mistake with \var{New\_Question} and  specified it as

\medskip
\begin{tla}
New_Question == /\ num < max_num_q + 1                         \* Mistake !!
                /\ new_question_enabled = TRUE                     
                /\ new_question_enabled' = FALSE
                /\ num' = num + 1
                /\ input_enabled' = TRUE
                /\ result' = ""
                /\ UNCHANGED <<count_right, count_wrong, check_enabled>>     
\end{tla}
\begin{tlatex}
 \@x{ New\_Question \.{\defeq} \.{\land} num \.{<} max\_num\_q \.{+}
 1\@s{98.39}}%
\@y{%
  Mistake !!
}%
\@xx{}%
\@x{\@s{82.75} \.{\land} new\_question\_enabled \.{=} {\TRUE}}%
\@x{\@s{82.75} \.{\land} new\_question\_enabled \.{'} \.{=} {\FALSE}}%
\@x{\@s{82.75} \.{\land} num \.{'} \.{=} num \.{+} 1}%
\@x{\@s{82.75} \.{\land} input\_enabled \.{'} \.{=} {\TRUE}}%
\@x{\@s{82.75} \.{\land} result \.{'} \.{=}\@w{}}%
 \@x{\@s{82.75} \.{\land} {\UNCHANGED} {\langle} count\_right ,\, count\_wrong
 ,\, check\_enabled {\rangle}}%
\end{tlatex}

\medskip
\noindent
then TLC would report a
deadlock: In this design, it is possible to do \var{New\_Question} when
\var{num} is \var{max\_num\_q}. Then, after this action, \var{num} will be
incremented to $\var{max\_num\_q} + 1$.  Therefore, the precondition of
\var{Terminating} is violated, the system is deadlocked.

\section{Related Work}
\label{sec:related}

Ever since the emergence of web applications, their formal analysis has been an
active research field. Model checking has been recognized as helpful.
Research so far mainly focuses on the analysis of
explicit navigation structures (hypertext-base applications with explicit
links). For an overview, see~\cite{alzahrani:phd:2015}.

Sylvain Hall{\'{e} et al.~\cite{halle-et-al:ase:2010} propose a method to apply
model checking to find potential navigation errors caused by browser
functionalities such as bookmarks or ``back'' button.  Deutsch et
al.~\cite{deutsch-sui-vianu:jcss:2007} discuss formal verification of
WebML~\cite{ceri_at_al:book:2002} models of data-driven web applications using
Abstract State Machines (ASM$^{+}$,~\cite{spielmann:phd:2000}). Miao and
Zeng~\cite{miao-zeng:iceccs:2007} consider a design model and an implementation
model of a web application, generate properties from the former, and 
check with SMV~\cite{berard_et_al:book:2001} if the latter does has the properties.  
Knapp and
Zhang~\cite{knapp-zhang:modellierung:2006} present a proposal of integrating
models of UWE~\cite{koch-et-al:2007} into a UML state
machine~\cite{omg:uml-2.5:2015} and model check it.

Compared to these methods, our approach uses \ta{} as the specification language
and its model checker TLC for model verification. More importantly, while the
aforementioned methods focus on hypertext-based web applications with
explicit navigation structures, our approach aims at documenting and
analyzing the workflows of \spa{}s that are defined implicitly by states of the
widgets. In particular, specifying \spa{}s with \ta{} makes it possible to
theorem-prove their properties, which will be part of our future work.

\section{Conclusions and Future Work}
\label{sec:conclusions}

We presented a practice friendly and easy-to-use approach to specifying
\spa{}'s workflows with \ta{}.  This way, the informal design of \spa{}s
can be documented formally, their implicit workflows make explicit.  Our
approach helps to understand designs of \spa{}s better, as well as to find
potential flaws in the design. 

Future work is possible in several directions. First, we plan to extend our
approach to handle more \spa{} features, such as asynchronous communication or
server activities.

Moreover, model checking is only amenable to finite-state systems. We plan to
extend your research and to use theorem
proving, which is also possible in \ta. Then it will be possible
to analyse infinite-state applications, and we could verify that if we removed the
constraint of maximum number of questions, then our math example would run
forever.
 
Last but not least, we plan to extend our research~\cite{zhang-zhao:seke:2018} of
analysing AngularJS~\cite{angularjs:2018} applications to generate the \ta{}
specification automatically from code. This would greatly improve our abilities
to analyse running AngularJS code. 

\bibliographystyle{plain}
\bibliography{bib}

\begin{thebibliography}{10}

\bibitem{alzahrani:phd:2015}
Mohammed~Yahya Alzahrani.
\newblock {\em {Model Checking Web Applications}}.
\newblock PhD thesis, Heriot-Watt University, 2015.

\bibitem{berard_et_al:book:2001}
B{\'{e}}atrice B{\'{e}}rard, Michel Bidoit, Alain Finkel, Fran{\c{c}}ois
  Laroussinie, Antoine Petit, Laure Petrucci, Philippe Schnoebelen, and Pierre
  McKenzie.
\newblock {\em {Systems and Software Verification, Model-Checking Techniques
  and Tools}}, chapter~12.
\newblock Springer, 2001.

\bibitem{ceri_at_al:book:2002}
Stefano Ceri, Piero Fraternali, Aldo Bongio, Marco Brambilla, Sara Comai, and
  Maristella Matera.
\newblock {\em {Designing Data-Intensive Web Applications}}.
\newblock Morgan Kaufmann, 2002.

\bibitem{clarke_et_al:2018}
Edmund~M. Clarke, Thomas~A. Henzinger, Helmut Veith, and Roderick Bloem,
  editors.
\newblock {\em {Handbook of Model Checking}}.
\newblock Springer, 2018.

\bibitem{deutsch-sui-vianu:jcss:2007}
Alin Deutsch, Liying Sui, and Victor Vianu.
\newblock {Specification and verification of data-driven Web applications}.
\newblock {\em J. Computer and System Sciences}, 73(3):442--474, 2007.

\bibitem{gallier:2018}
Jean~H. Gallier.
\newblock {\em {Logic for Computer Science: Foundations of Automatic Theorem
  Proving}}.
\newblock Dover Publications, 2018.

\bibitem{angularjs:2018}
Google.
\newblock Angularjs.
\newblock \url{https://angularjs.org/}, 2018.
\newblock Accessed on 2020-02-01.

\bibitem{halle-et-al:ase:2010}
Sylvain Hall{\'{e}}, Taylor Ettema, Chris Bunch, and Tevfik Bultan.
\newblock {Eliminating Navigation Errors in Web Applications via Model Checking
  and Runtime Enforcement of Navigation State Machines}.
\newblock In Charles Pecheur, Jamie Andrews, and Elisabetta~Di Nitto, editors,
  {\em Proc. 25$^\text{th}$ Int. Conf. Automated Software Engineering
  (ASE'10)}, pages 235--244. {ACM}, 2010.

\bibitem{knapp-zhang:modellierung:2006}
Alexander Knapp and Gefei Zhang.
\newblock {Model Transformations for Integrating and Validating Web Application
  Models}.
\newblock In Heinrich~C. Mayr and Ruth Breu, editors, {\em Proc. Modellierung
  2006 (MOD'06)}, volume P-82 of {\em Lect. Notes Informatics}, pages 115--128.
  Gesellschaft f{\"u}r Informatik, 2006.

\bibitem{koch-et-al:2007}
Nora Koch, Alexander Knapp, Gefei Zhang, and Hubert Baumeister.
\newblock {UML-Based Web Engineering: An Approach Based on Standards}.
\newblock In Luis Olsina, Oscar Pastor, Gustavo Rossi, and Daniel Schwabe,
  editors, {\em Web Engineering: Modelling and Implementing Web Applications},
  volume~12 of {\em Human-Computer Interaction Series}, chapter~7, pages
  157--191. Springer-Verlag, 2007.

\bibitem{kroeger-merz:2008}
Fred Kr{\"{o}}ger and Stephan Merz.
\newblock {\em {Temporal Logic and State Systems}}.
\newblock Texts in Theoretical Computer Science. An {EATCS} Series. Springer,
  2008.

\bibitem{lamport:2003}
Leslie Lamport.
\newblock {\em {The TLA+ Language and Tools for Hardware and Software
  Engineers}}.
\newblock Addison-Wesley, 2003.

\bibitem{miao-zeng:iceccs:2007}
Huaikou Miao and Hongwei Zeng.
\newblock {Model Checking-based Verification of Web Application}.
\newblock In {\em Proc. 12$^{\text{th}}$ Int. Conf. Engineering Complex
  Computer Systems (ICECCS'07)}, pages 47--55. {IEEE} Computer Society, 2007.

\bibitem{omg:uml-2.5:2015}
OMG.
\newblock {Unified Modeling Language, Version 2.5}.
\newblock Specification, Object Management Group, 2015.
\newblock \url{http://www.omg.org/spec/UML/2.5/PDF/}, Accessed on 2020-02-02.

\bibitem{spielmann:phd:2000}
Marc Spielmann.
\newblock {\em {Abstract state machines: verification problems and
  complexity}}.
\newblock PhD thesis, Rheinisch-Westf\"alische Technische Hochschule Aachen,
  2000.

\bibitem{wikipedia:spa}
Wikipedia.
\newblock {Single-page Application}.
\newblock
  \url{https://en.wikipedia.org/w/index.php?title=Single-page_application&oldid=938116540},
  2020.
\newblock Accessed: 2020-02-01.

\bibitem{zhang-zhao:seke:2018}
Gefei Zhang and Jianjun Zhao.
\newblock {Visualizing Interactions in AngularJS-based Single Page Web
  Applications}.
\newblock In {\em Proc. 30$^\text{th}$ Int. Conf. Software Engineering \&
  Knowledge Engineering (SEKE'18)}, pages 403--408. KSI Research Inc., 2018.

\end{thebibliography}

\end{document}